\documentclass[aps,twocolumn,preprintnumbers, amsmath, amssymb, floatfix]{revtex4}
\usepackage{dcolumn}
\usepackage{bm}
\usepackage{graphicx}
\def\be{\begin{equation}}
\def\ee{\end{equation}}
\def\beq{\begin{eqnarray}}
\def\eeq{\end{eqnarray}}

\def\bay{\begin{array}}
\def\eay{\end{array}}

\begin{document}
\preprint{smw05-sars-conf01}
\title{Progress with a Universal Theory of Relativity}

\author{Sanjay M. Wagh}
\affiliation{Central India Research Institute, \\
Post Box 606, Laxminagar, Nagpur 440 022, India}
\email{cirinag_ngp@sancharnet.in}

\date{October 25, 2005}
\begin{abstract}
In this presentation, I will summarize the present status of the
developments with a Universal Theory of Relativity \cite{smw-utr}.
Some general challenges to be overcome will also be discussed. \\

\centerline{Presented at the SARS Einstein Centennial Meeting,
September 25 - 26, 2005, Durban, South Africa}
\end{abstract}
\maketitle

\newpage

Newtonian physics is based on four basic conceptions, namely,
inertia, force, source of the force and a law of motion. Inertia is
the opposition of a physical body to a change in its state of motion
as perceived by an observer, the force is the ``cause'' behind the
motion of that physical body and, since only another physical body
can provide such a cause of the motion, every physical body is
attributed an appropriate ``source strength'' as per these
conceptions. Gravitational mass and electrostatic charge are two
such source characteristics attributed to physical bodies.

Due to the mutual {\it logical independencies\/} of these
conceptions, a law of force has to be {\em postulated\/} in
Newton's theory. Similarly, numerical values for inertia and
source strength(s) are also to be postulated or {\em specified by
hand\/} and an independent statement for the law of motion is also
needed in Newton's theory. It is only then that the newtonian laws
of motion predict the ``path'' of a physical body that is imagined
as a material point of the Euclidean space which underlies the
mathematical framework of Newton's theory. In this way, Newton's
mathematical framework makes a contact with the physical world.

The classical/newtonian field theory too postulates four, mutually
logically independent, conceptions, equivalent to the
aforementioned four, to account for the physical phenomena from
fluid dynamics, electromagnetism etc.

Due to the mutual logical independencies of involved conceptions,
newtonian theories cannot, as is well known, explain the origins
of inertia and the source characteristics (of forces) attributable
to physical matter. It is this same logical independency of
underlying conceptions that is also behind ``the equality of
inertia and gravitational mass of a physical body'' being an
assumption of Newton's theory. Still, newtonian theories explain a
large body of experimental results.

To explain the experimental results, we then require {\em four\/}
fundamental forces, {\em viz}, those of gravity, electromagnetism,
strong nuclear and weak nuclear interactions. Theoretical
physicists, in general, aim to explain these four interactions as
arising from a single entity, {\em ie}, to unify these four
forces. Clearly, the ``unification of the (fundamental) forces''
must of course provide an appropriate universal ``explanation''
which removes the mutual logical independencies of these involved
concepts \cite{smw-utr}. Evidently, this aim requires departures
from the newtonian framework.

The (orthodox) Quantum Theory replaces the concept of force with
that of a potential for the force. (But,  the source of potential
in quantum theory is the same as the source of the corresponding
newtonian force.) Schr\"{o}dinger's equation or Heisenberg's
operators of the quantum theory explicitly involve only the
potential generated by the ``source'' of the force. So is the
situation with the procedure of second quantization - the
creation, the annihilation and the number operators all involve
the potential for the force.

The  mathematical formalism of quantum theory leads us only to
probabilistic predictions about physical phenomena. Issues of
whether this probabilistic description is any ``complete''
description \cite{bohr2} of the physical reality or not are
``separate'' from those of the mutual logical independencies of
the underlying physical concepts.

Regardless of other issues, the inertia, the potential and the
``source'' for the potential are, once again, mutually independent
logical conceptions. Also, an independent statement of the law of
potential has to be postulated in quantum theory. Quantum theory
too is then based on four mutually logically independent concepts.

Therefore, (methods of) quantum theory also cannot explain the
origins of the inertia and the source characteristics (of
potentials or forces) attributable to physical matter. Then,
faithful applications \cite{aa-lqg, strings} of mathematical
methods of quantum theory cannot lead us to the ``true
unification'' of the four fundamental interactions.

That is to say, quantum theoretical methods ``postulate'' the
potentials corresponding to (basic) forces, but these methods
cannot provide a ``single'' explanation for those potentials,
their theoretical framework being too narrow.

Now, Einstein's principle of general relativity states
\cite{schlipp, subtle} that the laws of physics must be applicable
with respect to all the systems of reference \footnote{I will
reserve the word ``{\em Frame}'' for the mathematical notion to be
defined later and will use ``system of reference'' or ``reference
system'' to imply the use of a physical body as a reference for
the purpose of a physical measurement.}. This principle offers the
widest permissible conceptual basis for the entire physics.

To implement and also to test the usefulness of the ideas behind
this important principle, Einstein proposed his ``preliminary field
equations'' by replacing only the concept of gravitational force by
that of the curvature of the spacetime geometry while treating all
the other source characteristics of physical matter (or
equivalently, all the non-gravitational forces) as forming the
energy-momentum tensor.

Although encouraged by the ``initial successes'' of his field
equations, Einstein was never comfortable \cite{schlipp, subtle}
with the fact that his field equations of general relativity
provided only a theory of gravity. He never stated his (logical)
reasons explicitly, but the following illustrates the origins of
Einstein's unease with his field equations.

Consider a statement: {\em Every woman is replaced by a flower}.
Clearly, it is then logically inconsistent {\it to replace Queen
Cleopatra alone by a Ghost}. Needless to say, any statements about
some actions of Queen's Ghost and their consequences are {\em
logically unacceptable\/} to us, then.

Hence, the statement of a logically consistent theoretical
alternative to the newtonian framework must be that {\em the
newtonian force is being replaced by such and such conception,
universally applicable to every newtonian force}.

[Quantum theory, for example, replaces {\em every\/} force by its
potential and is logically consistent in this respect. Hence,
methods of quantum theory must be obtainable in a theory that
``unifies'' the fundamental interactions because the methods of
the former are ``universal'' in replacing the force by the
corresponding potential.

Moreover, since quantum theory makes ``probabilistic''
predictions, the relation of the quantum theory with the theory
that unifies the fundamental interactions can be expected
\cite{schlipp} to be very similar to that of the (usual)
statistical physics with the classical newtonian theory.]

Thus, replacing {\em only one force}, say, of gravity, by
curvature of geometry while considering all non-gravitational
interactions as forming the energy-momentum tensor is logically
unacceptable. But, this is what Einstein equations are. Therefore,
all solutions of Einstein equations are then like different
actions of Queen's ghost.

Hence, even if certain of the many solutions of Einstein's
equations were to ``equationally explain'' some observations of
the real world, say, the bending of light, precession of the
perihelion of Mercury, gravitational redshift, changes in the
period of a binary pulsar system, cosmology etc, the ``curvature
of geometry'' is not any ``logical reason'' for the involved
physical phenomena.

\begin{figure*}
\centering \includegraphics[width=3in, height=2in]{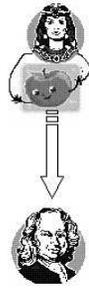}
\caption{\label{fig:ghost1} Ghost causing physical motion}
\end{figure*}

To state it bluntly, some action of Queen's Ghost cannot be any
acceptable ``reason'' for the falling of an apple on Newton's head
even if this action were to be always $1/r^2$. See Fig
(\ref{fig:ghost1}).

\begin{figure*}
\centering \includegraphics[width=4in,height=1.5in]{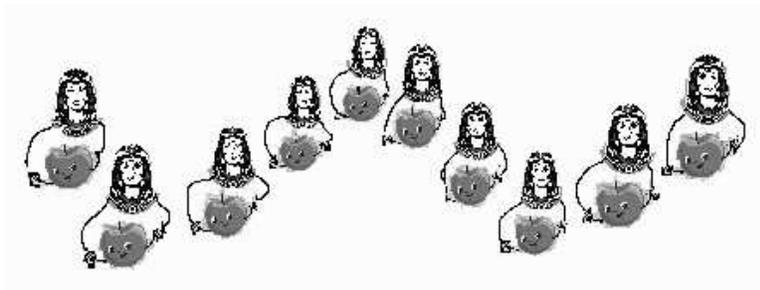}
\caption{\label{fig:ghost2} Ghost causing Wavy Motion}
\end{figure*}

Also, if Queen's ghost were to ``dance'' rhythmically holding the
hand of matter, say, in the LIGO detector then, some ``wavy''
motion of matter can be ``explained'' even accurately. See Fig
(\ref{fig:ghost2}). This however cannot form an explanation of the
wavy-motion of matter, even if such wavy motion is detected by any
detector \footnote{Changes in the curvature of 4-geometry
affecting LIGO matter are, then, like Queen's ghost changing the
location of LIGO matter in an undulatory fashion. {\bf\em We are
then led to challenge the ``physical existence of gravitational
radiation'' as ripples of curvature propagating on the spacetime
fabric.}}.

Now, the basis of Einstein's principle of general relativity is
the ``equality'' of {\em all\/} systems of reference, in relative
acceleration or not, for the description of all physical
phenomenon.

\begin{figure*}
\centering \includegraphics[width=2in,height=2in]{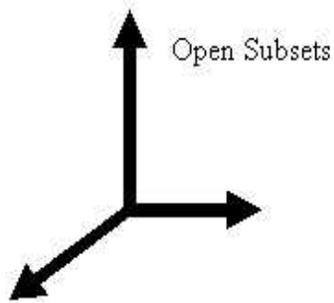}
\caption{\label{fig:system1} Physical construction of reference
system}
\end{figure*}

\begin{figure*}
\centering \includegraphics[width=4in,height=1.5in]{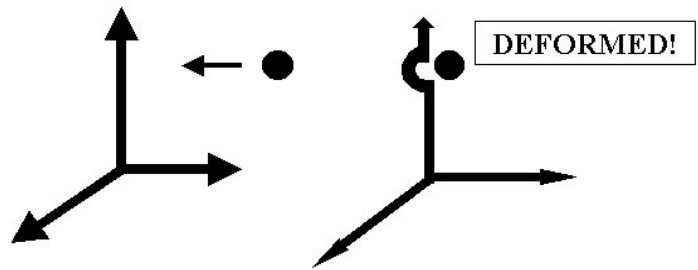}
\caption{\label{fig:system2} Change to the physical construction
of reference system as a physical phenomenon}
\end{figure*}

But, changes that can occur to the ``physical construction'' of
reference systems \cite{ein-pop} must also be the part of any such
description. See Fig (\ref{fig:system1}) and Fig
(\ref{fig:system2}). The equality of the description of physical
phenomena must then hold incorporating any possible physical
changes to the constructions of reference systems. This situation
necessitates ``changes'' also at the conceptual levels.

That is to say, the validity of the principle of general
relativity not only demands a mathematical framework that
incorporates the physical construction of reference systems but
also demands that we ``redefine'' various physical notions. I will
refer to this general theoretical framework \cite{smw-utr} as the
{\em Universal Theory of Relativity}.

In \cite{smw-utr}, I had discussed a mathematical framework of the
theories of measures and dynamical systems for this universal
relativity. However, it relied on certain physical observations
which did not, quite uniquely, fix the underlying mathematical
structures, although the aforementioned logical independencies
were removable in it. It could then be considered to be not
sufficiently general and, hence, not entirely satisfactory.

What, then, is the appropriate mathematical formalism for the
universal theory of relativity? It is the contention here that the
mathematical framework of Category Theory \cite{cat-0, cat-1} is
an appropriate basis for the universal relativity.

The rest of this paper constitutes a discussion about the relevant
mathematical notions \cite{banaschewski, macmor} and how these
notions could incorporate the ideas of universal relativity. Of
course, much more work is still needed on various fronts.

A partially ordered set (poset) $P$ is {\em complete\/} iff every
subset of $P$ has a lowest upper bound (lub or sup or join) and a
greatest lower bound (glb or inf or meet). A poset $P$ is then
complete iff, as a category, $P$ has all limits and all co-limits.

A {\em lattice\/} is a poset having, as a category, all binary
products and all binary co-products. A {\em complete lattice\/}
with an initial object 0 and a terminal object 1 is a complete
poset.

A {\em Frame\/} is a complete lattice in which binary meet
distributes over arbitrary joins. A {\em Frame homomorphism\/} is
a map between {\em Frames\/} preserving finite meets and arbitrary
joins. For any topological space $X$, the lattice of its open
sets, ${\cal O}X$, forms a {\em Frame}, ordered by set inclusion.

The correspondence $X \to {\cal O}X$ is functorial with the
functor ${\cal O}: \mathbb{T}op \to \mathbb{F}rm$ being a
contravariant functor, with $\mathbb{T}op$ being the category of
topological spaces and $\mathbb{F}rm$ being the category of {\em
Frames}. Any continuous map $f: X \to Y$ determines a {\em Frame
homomorphism\/} ${\cal O}f: {\cal O}Y \to {\cal O}X$ with ${\cal
O}f(U) = f^{-\,1}(U)$ for all $U \in {\cal O}Y$.

The {\em Spectrum functor\/} in the opposite direction $\Sigma:
\mathbb{F}rm \to \mathbb{T}op$ assigns to each {\em Frame\/} $L$
its {\em spectrum\/} $\Sigma L$ that is the space of all
homomorphisms $\xi: L \to \mathbf{2}$, $\mathbf{2}$ being the
2-element lattice, with each homomorphism being called a {\em
point of $L$}, with open sets $\Sigma_a= \left\{ \xi \in \Sigma L
| \xi{a} = 1 \right\}$ for any $a\in L$. It also assigns to each
of the {\em Frame homomorphisms\/} $h: M \to L$ a continuous map
$\Sigma h : \Sigma L \to \Sigma M$ such that $\Sigma h(\xi) = \xi
h$ with $(\Sigma h)^{-\,1}(\Sigma_a) = \Sigma_{h(a)}$ for any
$a\in M$.

The two functors ${\cal O}$ and $\Sigma$ are adjoint on the right
with the unit  of this adjunction being denoted by $\eta_L$  and
the co-unit by $\varepsilon_X$.

{\em Frames\/} for which $\eta_L$ is an isomorphism are called
{\em spatial}. ${\cal O}X$ is trivially a {\em spatial Frame}.

Spaces for which $\varepsilon_X$ is a homomorphism are called the
{\em Sober Spaces}. Category of Sober spaces $\mathbb{S}ob$ is
dually equivalent to the full subcategory of $\mathbb{S}p\,
\mathbb{F}rm$ of {\em spatial Frames}.

Category opposite to $\mathbb{F}rm$ is called as the category
$\mathbb{L}oc$ of {\em locales\/} and it contains $\mathbb{S}ob$
as a full subcategory. Category $\mathbb{L}oc$ is often used in
the relevant mathematical literature on pointless topology which
provides a wider basis than the usual topological structure.

Next, a {\em sheaf\/} is \cite{macmor} a ``continuous set-valued''
function. A sub-sheaf of a sheaf $F$ over a set $X$ is a
sub-functor of $F$ which is itself a sheaf. Sheaves over a set $X$
form the category $\mathbb{S}h(X)$. Categorical co-limits of
finite limits in $\mathbb{S}h(X)$ provide {\em geometric
constructions\/} which are point-wise and providing sheaves.
Category $\mathbb{S}h(X)$ of sheaves over $X$ then contains an
object - called the {\em generic set\/} - from which every other
object of $\mathbb{S}h(X)$ can be geometrically constructed.

When $X$ is a topological space, sub-sheaves of the terminal
object of $\mathbb{S}h(X)$ provide the {\em open\/} subsets of
$X$. When $X$ are Sober spaces, continuous maps between them are
functors between the corresponding categories of sheaves
preserving the geometric constructions.

This above provides \cite{macmor, sdg} a generalization of a
topological space - the {\em generalized space of sets\/} -
because the generic set is, in general, {\em not\/} isomorphic to
the initial object of $\mathbb{S}h(X)$, and the sub-objects of the
terminal object of $\mathbb{S}h(X)$ not necessarily its elements.
Consequently, the notion of ``continuity'' acquires, for example,
a new meaning in terms of the generic set.

Topos Theory was then developed \cite{fwlaw-01, fwlaw-02,
fwlaw-03, macmor} based on a complete Cartesian closed category
with a sub-object classifier - a Topos. Also, the differential
structure built by giving up the Law of Excluded Middle in the
Euclidean setup leads to Synthetic Differential Geometry
\cite{sdg}.

Using a topos equipped with the notion of an infinitesimal
time-interval, it is then possible \cite{fwlaw-01, fwlaw-02,
fwlaw-03} to recast the newtonian framework into the topos
theoretic language. In \cite{kock-reyes}, a wave equation is also
discussed from this point of view.

Physically, the construction of a reference system must
necessarily use physical bodies. For its mathematical expression,
we may consider a physical body as a collection of open subsets of
some topological space $X$ \footnote{Objection that the coordinate
axes (intervals) be closed, compact and bounded, since such is the
case with the real line $\mathbb{R}$, does not apply here because
physically interesting quantity is only the measurable or
observable ``distance'' between bodies and is definable for {\em
Frames\/} as ``distance between sets''. I thank Prof B
Banaschewski for raising this issue during a short discussion.}.
Then, the physical construction of a reference system is
mathematically representable as a {\em Frame}.

With the above identification, the category $\mathbb{F}rm$ of {\em
Frames\/} or, equivalently, the category $\mathbb{L}oc$ of locales
could form the underlying mathematical basis for Universal
Relativity. The category $\mathbb{S}h(X)$ would then be equally
relevant. Issues regarding the physical characteristics of
material bodies will then involve suitable {\em measures over
Frames}. These issues have not been addressed as yet.

Furthermore, the category $\mathbb{F}rm$ is not a Topos. (There
exist many other categories which are not toposes.) Consequently,
it is unclear whether the Topos Theory could form a mathematical
basis for the universal relativity as it did \cite{fwlaw-01} for
the newtonian theoretical framework.

However, the notion of a category is very general indeed and,
hence, could form the basis for universal relativity. An important
unanswered mathematical issue is then of defining an appropriate
notion of (generalized) measures in the general setting of the
category theory.

Notably, very general ideas from Physics and Mathematics appear to
become quite identical here. Starting with the limited formalism
in \cite{smw-utr}, this is certainly a significant progress.

\newpage
\acknowledgments Discussions with Partha Ghosh as well as with
Gareth Amery, Sunil Maharaj and many others helped me focus my own
ideas. I thank them as well as all the organizers of the SARS
Einstein Centennial Meeting for excellent arrangements and for an
opportunity to present this work. I also wish to thank Dr T Smith
of IOP for pointing out to me the exact reference for the work in
\cite{aa-lqg}.

\goodbreak
\smallskip

\smallskip

\smallskip

\vfill

\end{document}